\documentclass[prc]{revtex4}

\usepackage{graphicx,dcolumn,array,bm,amsmath}

\begin{document}
\title{Systematic study of proton-neutron pairing correlations in the nuclear shell model}

\author{Y. Lei}

\address{Department of Physics, Shanghai Jiao Tong University, Shanghai, 200240, China;  Bartol Research Institute and Department of Physics and
Astronomy, University of Delaware, Newark, Delaware 19716, USA}

\author{S. Pittel}

\address{Bartol Research Institute and Department of Physics and
Astronomy, University of Delaware, Newark, Delaware 19716, USA}

\author{N. Sandulescu}

\address{Institute of Physics and Nuclear Engineering, 76900
Bucharest, Romania}

\author{A. Poves}

\address{Departamento de Fisica Teorica and IFT UAM/CSIC, Universidad Autonoma de Madrid,
28049, Madrid Spain.}

\author{B. Thakur}

\address{Bartol Research Institute and Department of Physics and
Astronomy, University of Delaware, Newark, Delaware 19716, USA; LONI
Institute and Center for Computational Technology, Louisiana State
University, Baton Rouge, LA 70803-4001 USA}

\author{Y.M. Zhao}

\address{Department of Physics, Shanghai Jiao Tong University, Shanghai, 200240, China}

\begin{abstract}
A shell-model study of proton-neutron pairing in $2p1f$ shell nuclei
using a parametrized hamiltonian that includes deformation and
spin-orbit effects as well as isoscalar and isovector pairing is
reported. By working in a shell-model framework we are able to
assess the role of the various modes of proton-neutron pairing in
the presence of nuclear deformation without violating symmetries.
Results are presented for $^{44}$Ti, $^{45}$Ti, $^{46}$Ti, $^{46}$V
and $^{48}$Cr to assess how proton-neutron pair correlations emerge
under different scenarios. We also study how the presence of a
one-body spin-obit interaction affects the contribution of the
various pairing modes.
\end{abstract}

\maketitle

\begin{center}
\end{center}

\section{Introduction}

It is generally believed that $proton-neutron$ ($pn$) pairing is
important in nuclei with roughly equal numbers of neutrons and
protons \cite{Goodman}. The standard technique for treating these
correlations is through the Bardeen Cooper Schrieffer (BCS) or
Hartree Fock Bogolyubov (HFB) approximation, generalized to include
the $pn$ pairing field in addition to the $nn$ and $pp$ pairing
fields \cite{Goodman}. Questions arise, however, as to whether these
methods can adequately represent the physics of the competing modes
of pair correlations, without full restoration of symmetries
\cite{DP}.

Important insight into this issue has been achieved recently in the
context of exactly-solvable models that include these different
pairing modes. Analysis of the SO(8) model \cite{so8}, in which
isoscalar and isovector pairing act in either a single active
orbital or a series of degenerate orbitals, suggests that isospin
restoration or equivalently quartet correlations are extremely
important, especially near $N=Z$ \cite{DP}. More recent studies,
carried out for models involving non-degenerate orbitals
\cite{Errea}, reinforce earlier conclusions as to where isoscalar
pairing correlations should be most important
\cite{Satula},\cite{Sandulescu}. Furthermore, they make possible the
description of deformation, as is critical for systems with $N
\approx Z$,  by treating the non-degenerate orbitals as
Nilsson-like. However, it is still not possible to restore certain
symmetries within these models, for example rotational symmetry.

As a consequence, there still remain many open issues concerning the
role of the different possible modes of pairing in $N \approx Z$
nuclei. In this work, we report a systematic study of pairing
correlations in the context of the nuclear shell model, whereby
deformation can be readily included and symmetries maintained throughout. In
this way, we are able to address many of the open issues on the role
of the various pairing modes in the presence of nuclear deformation.

An outline of the paper is as follows. In section II, we briefly
describe our model and then in section III describe selected
results. Finally, in section IV we summarize the key conclusions of
the paper.

\section{Our model}

To address in a systematic way the role of pairing correlations in
the presence of nuclear deformation, we consider neutrons and
protons restricted to the orbitals of the $2p1f$ shell outside a
doubly-magic $^{40}Ca$ core and interacting via a schematic
hamiltonian

\begin{equation}
H= \chi \left( Q \cdot Q + a P^{\dagger} \cdot P + b S^{\dagger}
\cdot S + \alpha \sum_i \vec{l}_i \cdot \vec{s}_i \right)
\label{hamiltonian}
\end{equation}
where $ Q = Q_n +Q_p$ is the mass quadrupole operator, $P^{\dagger}$
creates a correlated $L=0$, $S=1$, $J=1$, $T=0$ pair and
$S^{\dagger}$ creates a correlated $L=0$, $S=0$, $J=0$, $T=1$ pair.
The first term in the hamiltonian produces rotational collective
motion, whereas the second and third term are the isoscalar and
isovector pairing interactions, respectively, whose matrix elements
can be found in ref. (\cite{Poves}). The last term is the one-body
part of the spin-orbit interaction, which splits the $j=l \pm 1/2$
levels with a given $l$.

We carry out shell-model calculations systematically as a function
of the various strength parameters. We begin with pure SU(3)
rotational motion \cite{su3} associated with the $Q \cdot Q$
interaction and then gradually ramp up the various SU(3)-breaking
terms to assess how they affect the rotational properties. This
includes the isocalar and isovector pairing interactions and the
spin-orbit term.

We first consider the nucleus $^{44}Ti$, with $N_{\rm n}=N_{\rm
p}=2$, and then systematically increase $N_{\rm n}$ and $N_{\rm p}$
to study the role of the number of active neutrons and protons, e.g.
whether there is an excess of one over the other and whether the
nucleus is even-even, odd-mass or odd-odd. The nuclei we have
treated are $^{44}Ti$ ($N_{\rm n}=N_{\rm p}=2$), $^{45}$Ti ($N_{\rm
n}=2$, $N_{\rm p}=3$), $^{46}$Ti ($N_{\rm n}=2$, $N_{\rm p}=4$),
$^{46}$V ($N_{\rm n}=3$, $N_{\rm p}=3$), and $^{48}$Cr ($N_{\rm
n}=4$, $N_{\rm p}=4$). Some of the observables we have studied are
(1) the energies and associated BE(2) values of the lowest
rotational band, (2) the number of isovector $S^{\dagger}$ pairs and
(3) the number of isoscalar $P^{\dagger}$ pairs.  In the following,
we present selected results that derive from these calculations.

\section{Results}

\subsection{An optimal hamiltonian}

Before turning to our results for specific nuclei, we first ask
whether the hamiltonian (\ref{hamiltonian}) has sufficient
flexibility to realistically describe the nuclei under
investigation. Without making an effort towards an absolute fit, we
note that the choice $ \chi = -0.05 ~MeV,~a=b=12$, and $\alpha= 20$
gives an acceptable fit to the spectra of all the nuclei we have
considered.

\begin{figure}[tbp]
\begin{center}
\includegraphics[width=12cm]{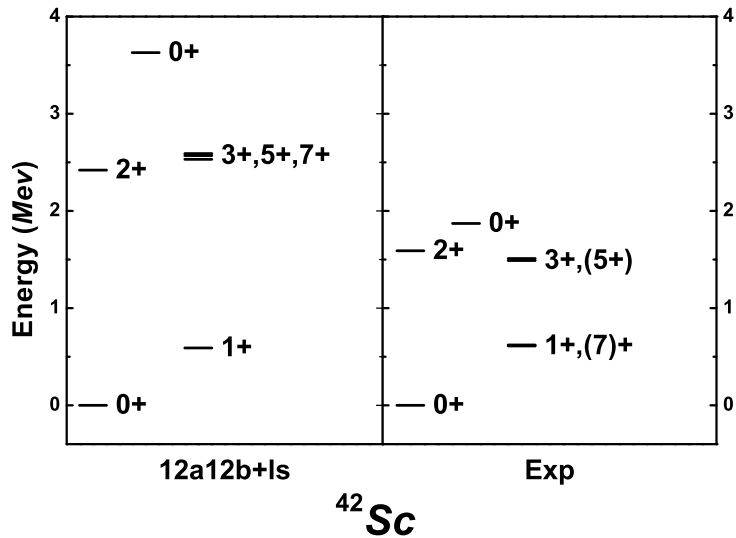}
\end{center}
\vspace{-1cm} \caption{\label{fig1}Comparison of experimental
spectra  for $^{42}Sc$ with the calculated spectra obtained using
the {\it optimal} hamiltonian described in the text. All energies
are in $MeV$.}
\end{figure}

We first show in figure 1 its prediction for $^{42}$Sc, in
comparison with the experimental spectrum. Other than its inability
to reproduce the low-lying $J^{\pi}=7^+$, $T=0$ state, the optimal
hamiltonian does an acceptable job in qualitatively describing the
features of the low-energy spectrum.  The lack of an acceptable
description of the $7^+$ state reflects the fact that our optimal
hamiltonian is missing the strong attraction between $f_{7/2}$
nucleons in the stretch configuration.

\begin{figure}[tbp]
\begin{center}
\includegraphics[width=14cm]{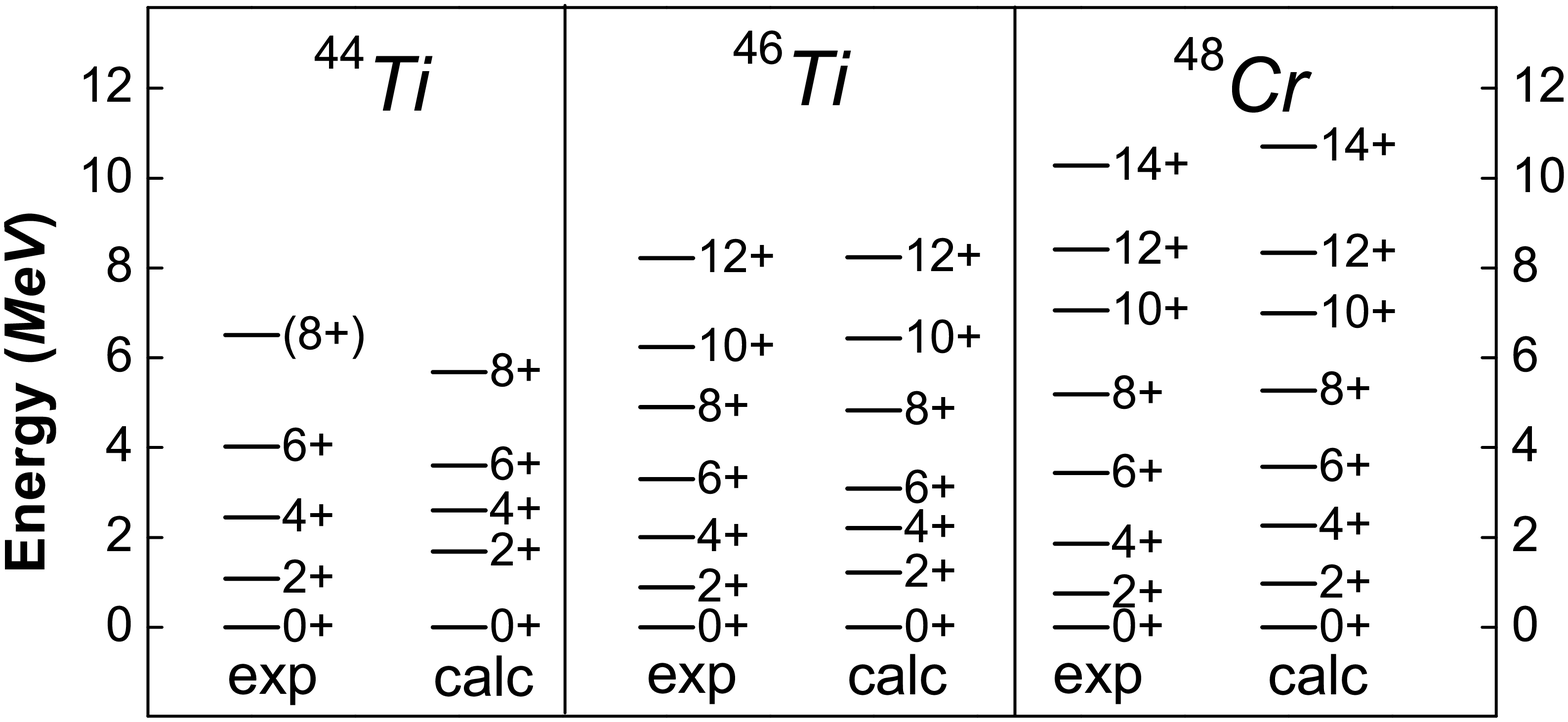}
\end{center}
\vspace{-1cm} \caption{\label{fig2}Comparison of experimental
spectra for $^{44}$Ti, $^{46}$Ti and $^{48}$Cr with the calculated
spectra obtained using the {\it optimal} hamiltonian described in
the text. All energies are in $MeV$.}
\end{figure}

In figure 2, we show its predictions for $^{44}$Ti, $^{46}$Ti and
$^{48}$Cr. As can be seen, the non-rotational character of $^{44}$Ti
at low angular momenta is reproduced by our calculations, as are the
highly rotational patterns seen experimentally for the heavier
nuclei. As we will see later, even the experimentally observed
backbend in $^{48}$Cr is acceptably reproduced with this
hamiltonian. We refer to the choice $a=b$ in the {\it optimal}
hamiltonian as the SU(4) choice, from the dynamical symmetry that
derives from this choice of parameters in the SO(8) model.

\subsection{$^{44}$Ti}

\begin{figure}[tbp]
\begin{center}
\includegraphics[width=18cm]{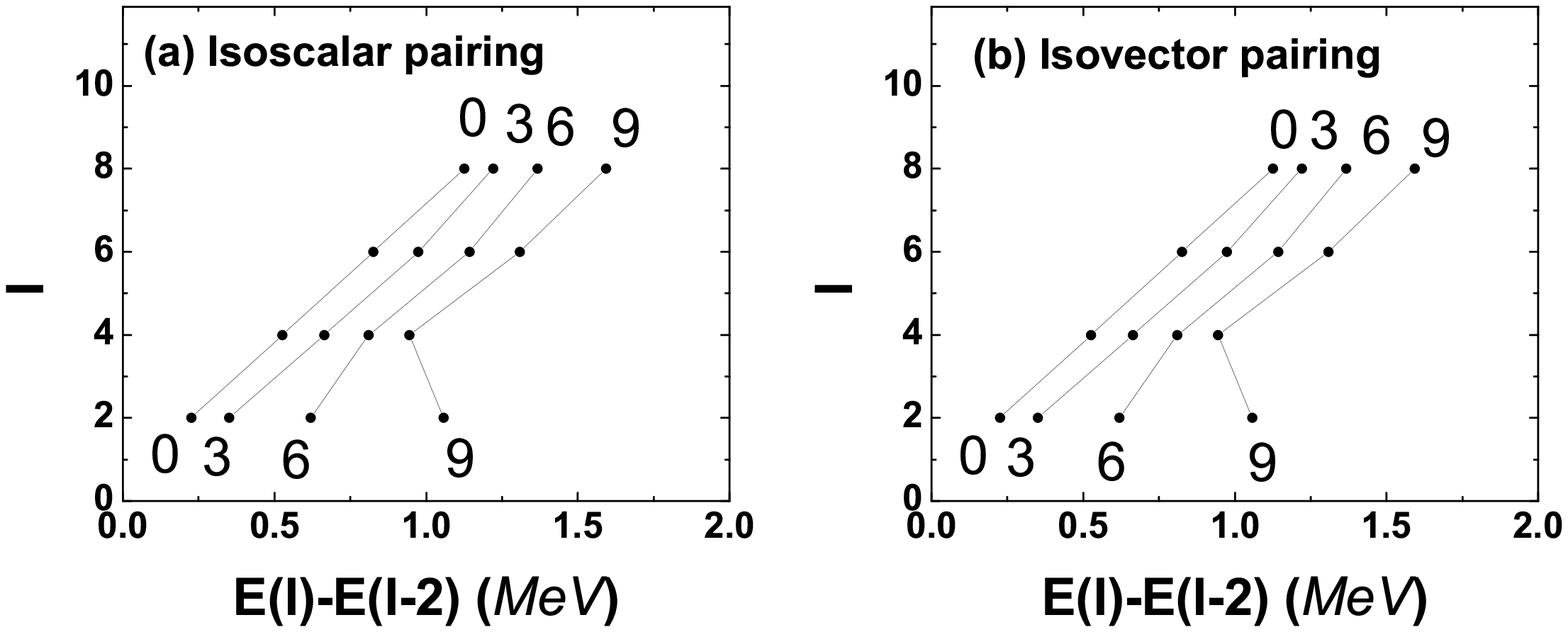}
\end{center}
\caption{\label{fig3} Calculated energy splittings $E(I)-E(I-2)$ (in
$MeV$) in the ground band of $^{44}$Ti  as a function of the
strength of the (a) isoscalar pairing interaction and the (b)
isovector pairing interaction, with no spin-orbit splitting. The
strengths of the respective pairing interactions are shown at the
ends of the lines, as they are elsewhere in the manuscript.}
\end{figure}

The first nucleus we discuss is $^{44}$Ti, with two active neutrons
and two active protons. In figure 3, we show the calculated energy
splittings $E(I)-E(I-2)$ associated with the ground-state band as a
function of the strength parameters $a$ and $b$ that define the
isoscalar and isovector pairing interactions, respectively. For
these calculations we assumed a quadrupole strength of
$\chi=-0.05~MeV$ {\it and no spin-orbit interaction}. As expected,
in the absence of a spin-orbit splitting the isoscalar and isovector
pairing interactions have precisely the same effect on the
properties of the ground state rotational band. The same conclusion
derives when we consider the effect of isoscalar and isovector
pairing on other observable properties when no spin-orbit splitting
is included. As an example, we show in figure 4 results for the
BE(2) values connecting the states in the ground band, again as a
function separately of the isovector and isoscalar pairing
strengths.

\begin{figure}[tbp]
\begin{center}
\hspace{5cm}
\includegraphics[width=16cm]{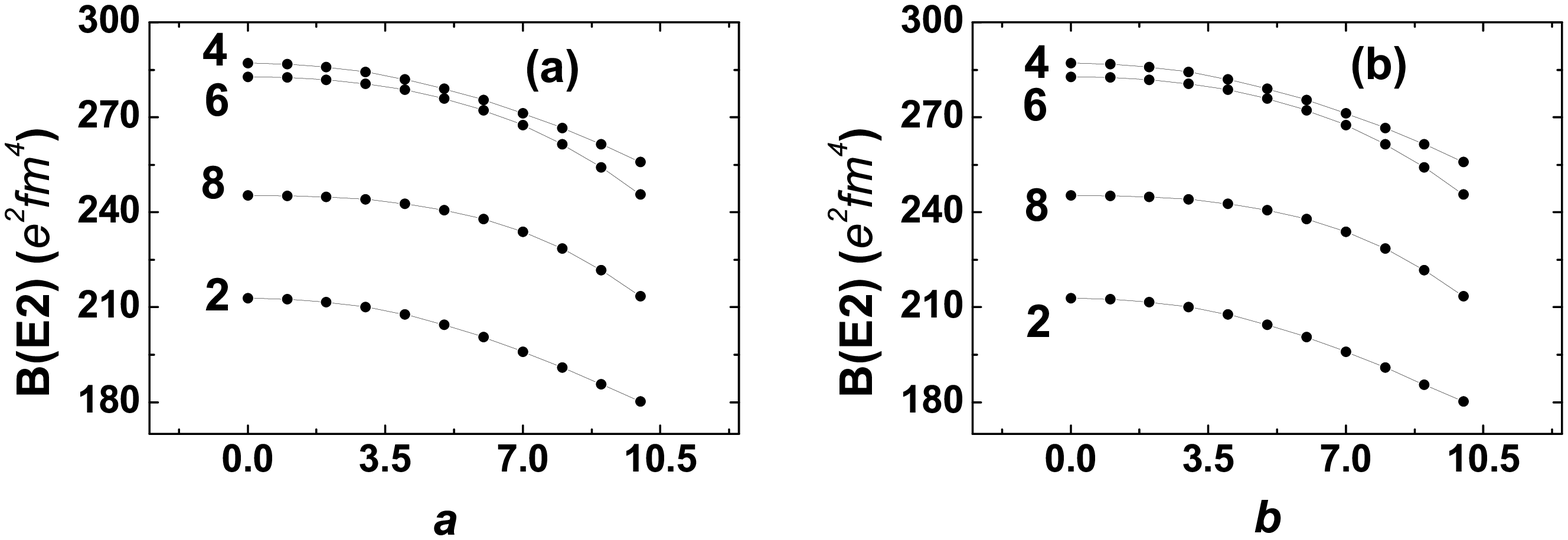}
\end{center}
\vspace{-1cm} \caption{\label{fig4} Calculated E2 transition matrix
elements $B(E2,I \rightarrow I-2)$ in the ground band of $^{44}$Ti
as a function of the strength of the (a) isoscalar pairing
interaction and of the (b) isovector pairing interaction , with no
spin-orbit splitting. The angular momentum $I$ of the initial state
appears to the left of each line. }
\end{figure}

We next show in figure 5 the same results as in figure 2, but now in
the presence of our realistic spin-orbit strength, $\alpha=20$. Here
we see that the effects of isoscalar and isovector pairing are very
different. In the presence of realistic single-particle energies,
isovector pairing produces a much more rapid loss of rotational
collectivity than isoscalar pairing.

\begin{figure}[tbp]
\begin{center}
\includegraphics[width=18cm]{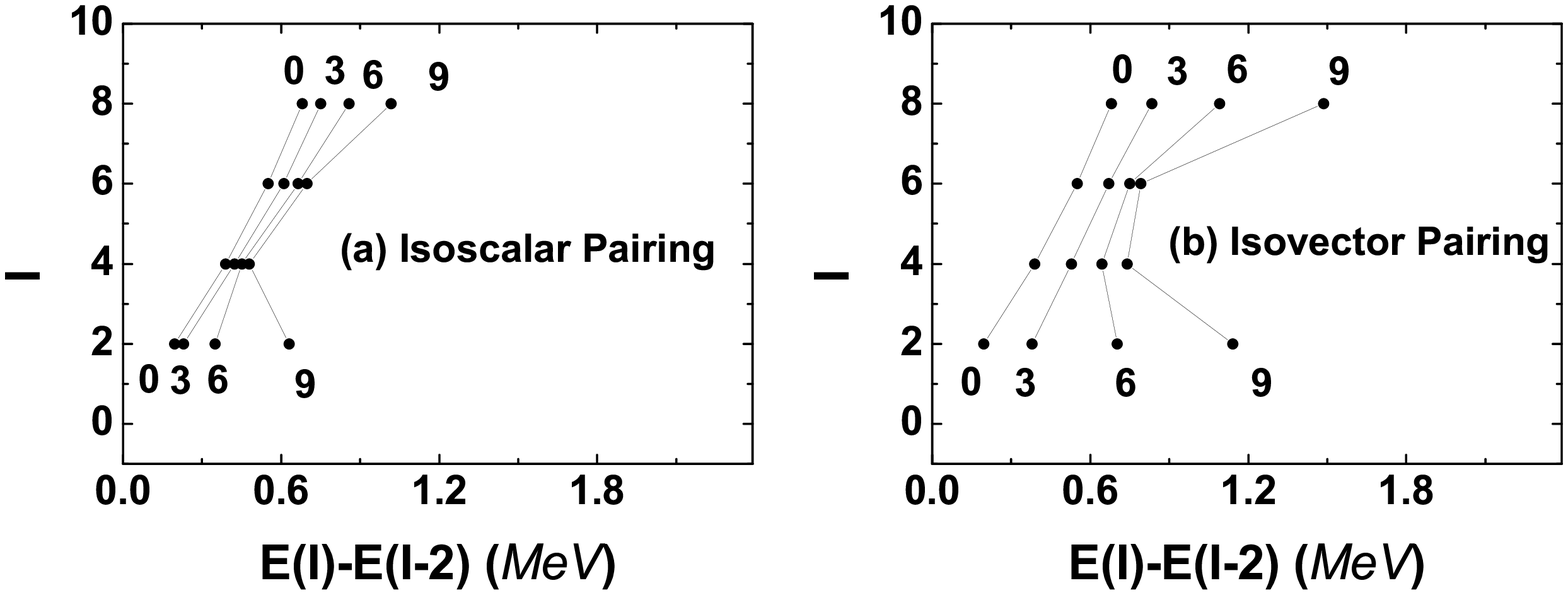}
\end{center}
\caption{\label{fig5} Calculated energy splittings $E(I)-E(I-2)$ in
$MeV$ in the ground band of $^{44}$Ti as a function of the strength
of the (a) isoscalar pairing interaction  and of the (b) isovector
pairing interaction, with the optimal spin-orbit term present. }
\end{figure}

Next we consider the relative effect of the spin-orbit interaction
on isovector and isoscalar pairing properties in the optimal SU(4)
limit, where both modes of pairing contribute with the same
strength. This is addressed in figure 6, where we show the average
number of $S^{\dagger}$ and $P^{\dagger}$ pairs in the ground band
as a function of the strength of the spin-orbit force. These are
determined by considering  $\langle S^{\dagger}\cdot S \rangle $ and
$\langle P^{\dagger}\cdot P \rangle $ and scaling them with respect
to the results that would derive from pure $T=0$ and $T=1$ pairing
hamiltonians (for a system of two pairs in an $\Omega=10$ shell),
respectively. While the number of isovector pairs does not change
substantially with increasing spin-orbit strength, the isoscalar
pair number is reduced dramatically, especially for the lower
angular momentum states of the band. We conclude, therefore, that
the spin-orbit interaction suppresses isoscalar pairing, already at
$N=Z$.  The mechanism whereby this suppression takes place was
discussed recently in ref. \cite{Bertsch}.

\begin{figure}[tbp]
\begin{center}
\includegraphics[width=18cm]{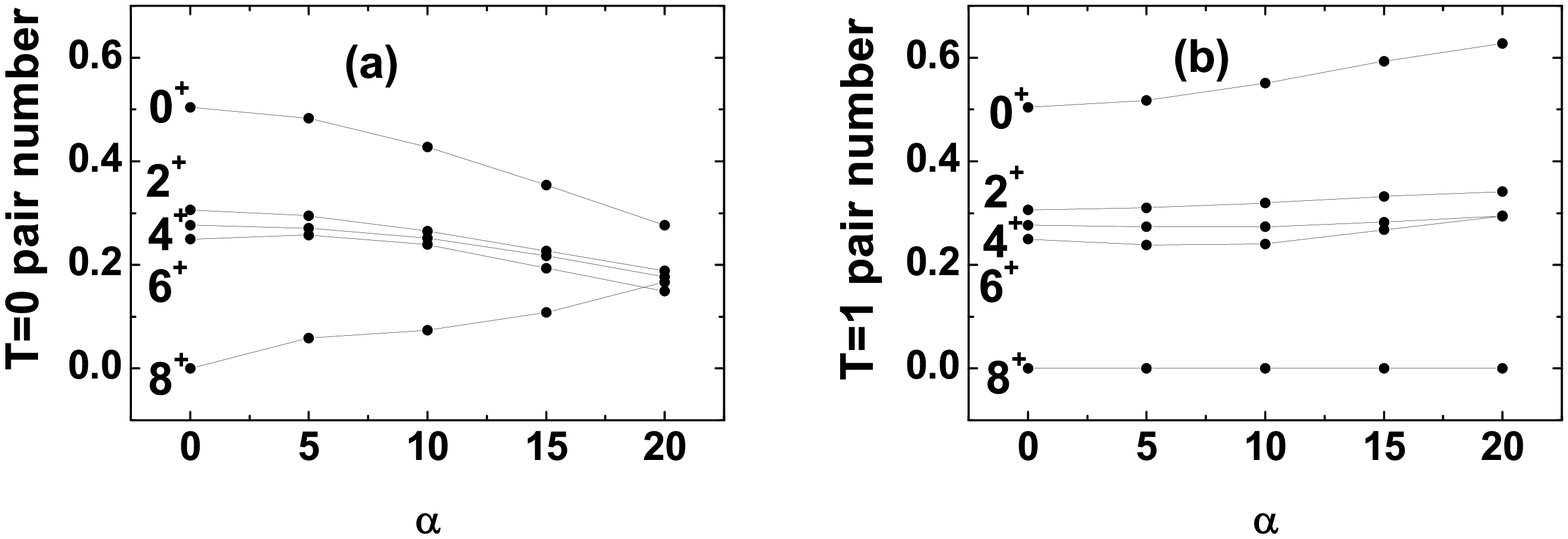}
\end{center}
\caption{\label{fig6} The number of (a) isocalar $P^{\dagger} $
pairs and (b) isovector $S^{\dagger}$  pairs in $^{44}$Ti as a
function of the strength of the spin-orbit interaction $\alpha$. All
other hamiltonian parameters are the optimal values. }
\end{figure}

\begin{figure}
\begin{center}
\includegraphics[width=12cm]{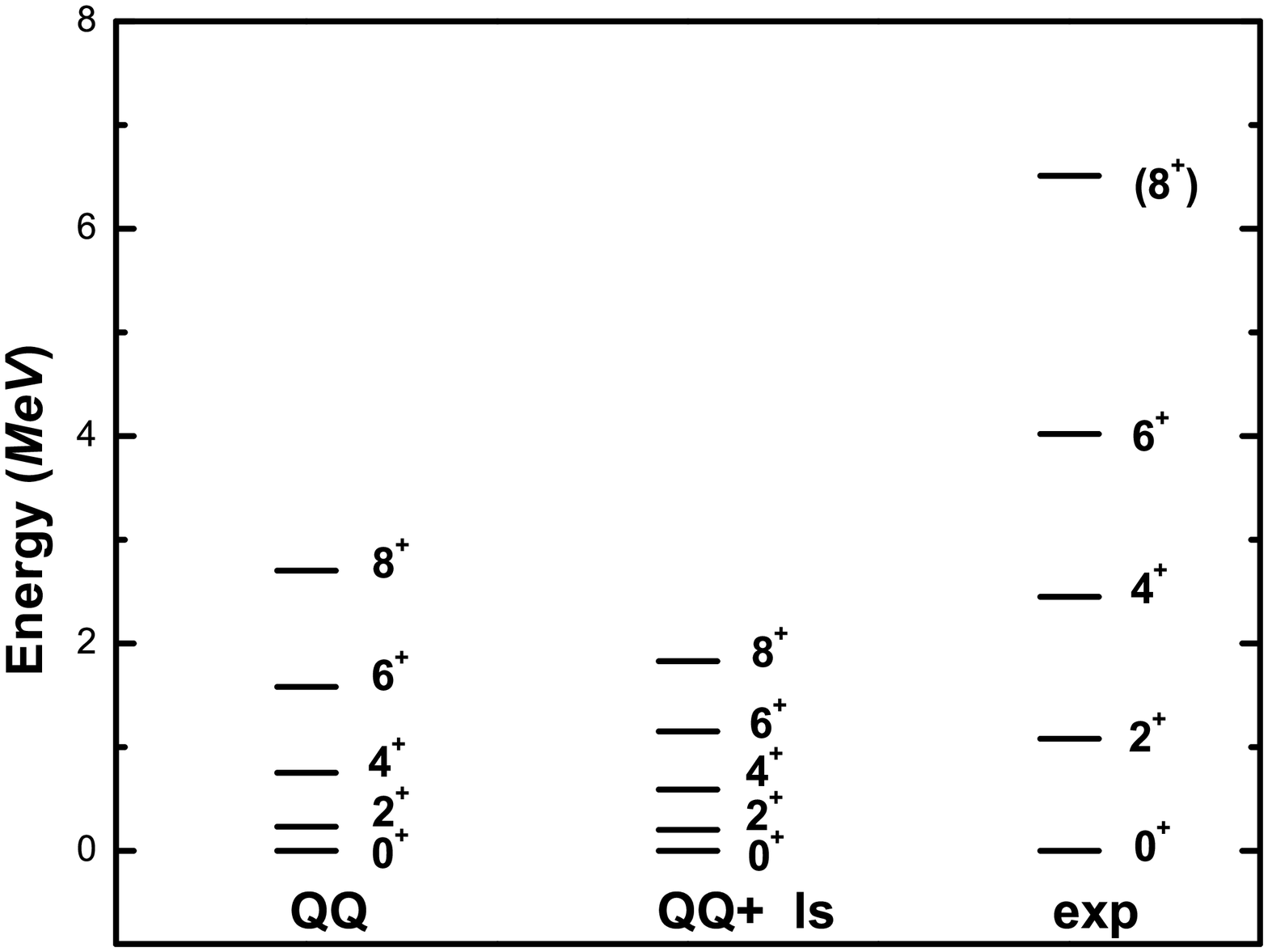}
\end{center}
\caption{\label{fig7}  Comparison of the experimental spectrum of
$^{44}$Ti with those obtained using a pure $Q \cdot Q$ interaction
and both a $Q \cdot Q$ interaction and a spin-orbit term. }
\end{figure}

Finally, in figure 7 we show the spectrum of $^{44}$Ti that derives
{\it solely} from turning on a strong spin-orbit force, i.e. with no
pairing present. We see that the spectrum is still highly
rotational, despite the fact that the resulting single-particle
energies are no longer SU(3)-like. To obtain the physical spectrum
with a non-rotational character, it is thus essential to have
pairing. It has been traditionally thought that it is the non-SU(3)
order of the single-particle levels that is responsible for the
non-rotational character seen in the experimental spectrum
\cite{Bhatt}, a conclusion that is not supported by our results. It
is pairing that is responsible for the non-rotational character of
$^{44}$Ti. This point was already suggested in Fig. 3, where we we
saw that for the physical pairing strengths a highly non-rotational
spectrum near the ground band emerged even in the absence of a
spin-orbit splitting. Further understanding of this conclusion
follows from the important role of quasi-SU(3) \cite{quasi-su3} in
producing deformation. Even though the spin-orbit interaction breaks
the SU(3) symmetry, it leaves quasi-SU(3) symmetry approximately
preserved.

\begin{figure}[tbp]
\begin{center}
\includegraphics[width=16cm]{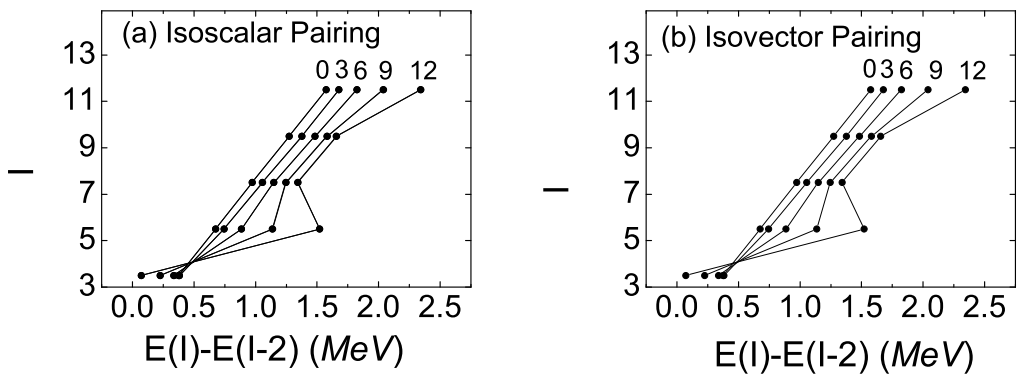}
\end{center}
\vspace{-2cm} \caption{\label{fig8}  Calculated energy splittings
$E_{I}-E_{I-2}$ in $MeV$ within the odd group of levels (see text)
of the ground state band of $^{45}$Ti as a function of the strength
of the (a) isoscalar pairing interaction  and of the (b) isovector
pairing interaction, with no spin-orbit term present. }
\end{figure}

\begin{figure}[tbp]
\begin{center}
\includegraphics[width=16cm]{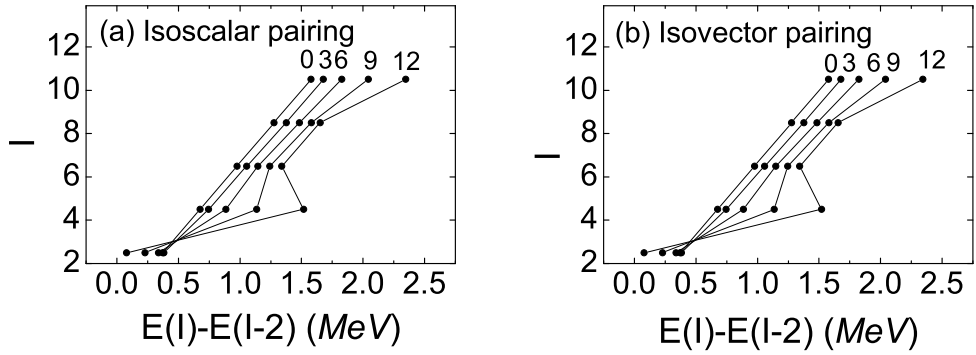}
\end{center}
\vspace{-2cm} \caption{\label{fig9} Calculated energy splittings
$E_{I}-E_{I-2}$ within the even group of levels (see text) of the
ground state band of $^{45}$Ti as a function of the strength of the
isoscalar pairing interaction (left panel) and of the isovector
pairing interaction (right panel), with no spin-orbit term present.
}
\end{figure}

\subsection{$^{45}$Ti}

Next we consider the odd-mass nucleus $^{45}$Ti, with one additional
neutron relative to $^{44}$Ti. For notational purposes, we divide
the results according to whether $I-1/2$ is odd (referring to this
as the odd group) or whether it is even (the even group). This
reflects the fact that states within the ground state band decay by
strong E2 transitions within their own groups.

In figures 8 and 9, we present results for the calculated energy
splittings of the ground band in the odd and even groups,
respectively, as a function of pure isoscalar (panel a) and pure
isovector (panel b) pairing, in both cases with no spin-orbit force
present. The results suggest that blocking due to an odd neutron
does not affect the symmetry between isoscalar and isovector pairing
when there is no spin-orbit force.

In figures 10 and 11, we show the corresponding results in the presence of a spin-orbit force,
with the optimal strength $\alpha=20$. Now isoscalar pairing is suppressed and there is a much
more dramatic effect of isovector pairing.

\begin{figure}[tbp]
\begin{center}
\includegraphics[width=16cm]{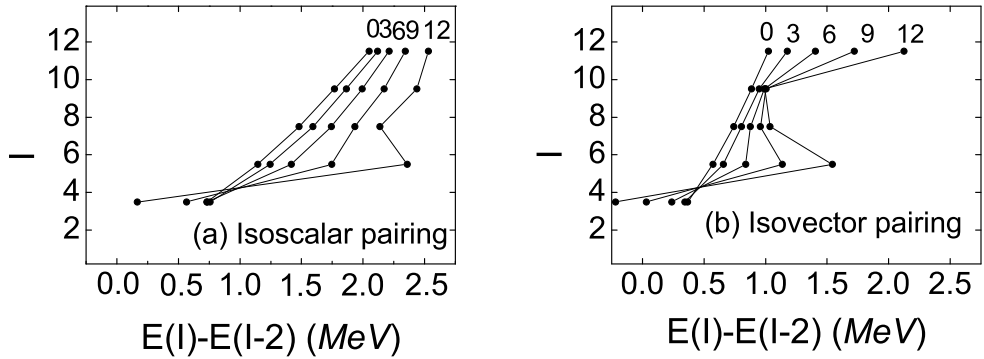}
\end{center}
\vspace{-1cm} \caption{\label{fig10}  Calculated energy splittings
$E(I)-E(I-2)$ in $MeV$  within the odd group of levels (see text) of
the ground state band of $^{45}$Ti as a function of the strength of
the (a) isoscalar pairing interaction  and of the (b) isovector
pairing interaction, with the optimal spin-orbit term present. }
\end{figure}

\begin{figure}[tbp]
\begin{center}
\includegraphics[width=16cm]{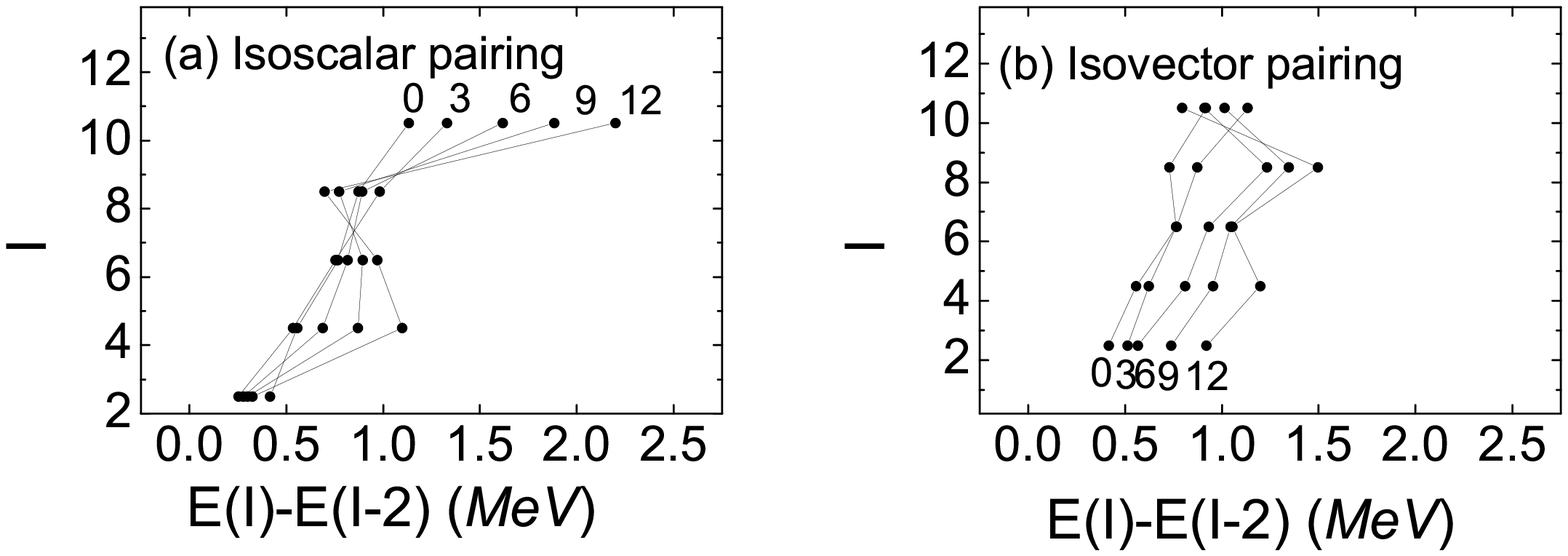}
\end{center}
\vspace{-1cm} \caption{\label{fi11} Calculated energy splittings
$E(I)-E(I-2)$ in $MeV$  within the even group of levels (see text)
of the ground state band of $^{45}$Ti as a function of the strength
of the (a) isoscalar pairing interaction and of the (b) isovector
pairing interaction, with the optimal spin-orbit term present. }
\end{figure}

\subsection{$^{46}$Ti}

\begin{figure}[tbp]
\begin{center}
\includegraphics[width=16cm]{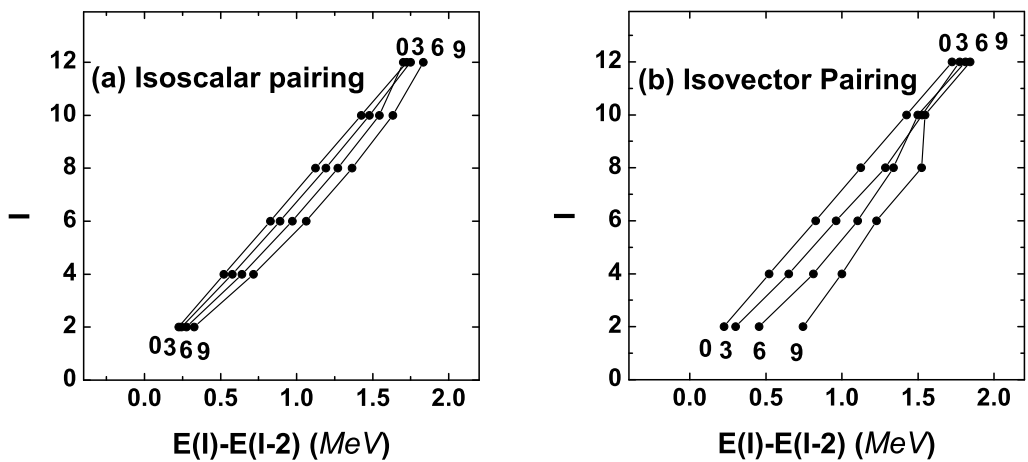}
\end{center}
\vspace{-1cm} \caption{\label{fig12}  Calculated energy splittings
$E(I)-E(I-2)$ in $MeV$ in the ground band of $^{46}$Ti as a function
of the strength of the (a) isoscalar pairing interaction and of the
(b) isovector pairing interaction, with no spin-orbit term present.
}
\end{figure}

Next we turn to $^{46}$Ti with two excess neutrons present. Here too
we compare the effect of the isoscalar and isovector pairing
interactions on deformation, showing the results in figure 12 with
no spin-orbit term present. Here the effect of isoscalar pairing is
strongly suppressed relative to isovector pairing, suggesting that
even without a spin-orbit term isoscalar pairing is very strongly
focused on those nuclei with $N=Z$ with a slight excess being
sufficient to suppress this pairing mode.

\subsection{$^{46}$V}

We next consider $^{46}$V, an odd-odd $N=Z$ nucleus. The $T=1$
states of $^{46}$V are of course precisely the same as those already
considered in $^{46}$Ti.

We first address in figure 13 the manner whereby the symmetry
between isocalar and isovector pairing in the absence of a
spin-orbit force is reflected in $^{46}$V. In the absence of
isoscalar and isovector pairing, the $J=1^+$ state and the $J=0^+$
state form a degenerate ground state doublet. When only isoscalar
pairing is turned on (panel a), the $J=1^+$ state is pushed down
below the $J=0^+$ state. When only isovector pairing is turned on
(panel b) the reverse happens and the $J=0^+$ is pushed down and
becomes the ground state. In the SU(4) limit (panel c) with equal
isovector and isocalar pairing strengths, the degeneracy reappears.

In figure 14, we show what happens in the presence of the physical
spin-orbit interaction, for equal isovector and isoscalar pairing.
Now the degeneracy is broken and the $0^+$ state emerges as the
ground state, as in experiment. The experimental splitting is 1.23
$MeV$, whereas our optimal hamiltonian produces a splitting of 1.05
$MeV$.

We should note that the first excited state in $^{46}$V
experimentally is found to be a $3^+$ state, at 801~$keV$. In our
calculations the lowest $3^+$ state occurs at significantly higher
energy, at 1.89~$MeV$. This may be related to the schematic nature
of our hamiltonian.

\begin{figure}
\begin{center}
\includegraphics[width=17cm]{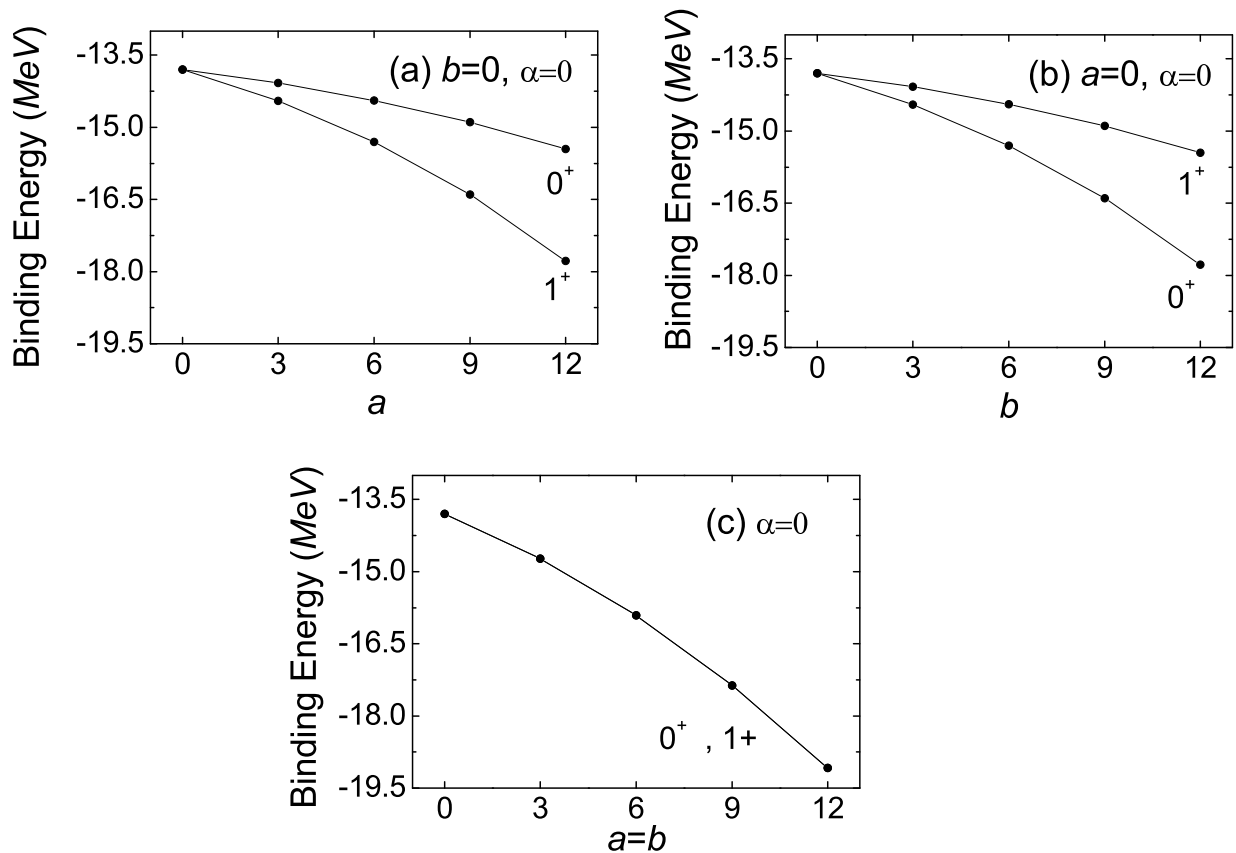}
\end{center}
\caption{\label{fig13}  Calculated energies in $MeV$ of the lowest
$J^{\pi}=0^+$ and $J^{\pi}=1^+$ states of $^{46}$V with no
spin-orbit term present.  Panel (a) shows the results of pure
isoscalar pairing, panel (b) shows the results of pure isovector
pairing and panel (c) shows the results of SU(4) pairing. }
\end{figure}

\begin{figure}
\begin{center}
\includegraphics[width=11cm]{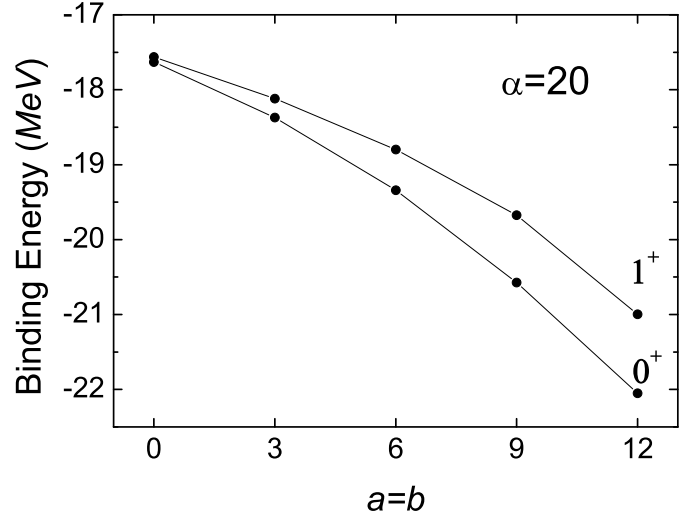}
\end{center}
\caption{\label{fig14}  Calculated energies in $MeV$ of the lowest
$J^{\pi}=0^+$ and $J^{\pi}=1^+$ states of $^{46}$V as a function of
the equal strength of  isoscalar and isovector pairing, with the
optimal spin-orbit term ($\alpha=20$) present.  }
\end{figure}

\subsection{$^{48}$Cr}

Lastly, we turn to $^{48}$Cr, which again has $N=Z$, but now with
two quartet structures present. Here we assume as our starting point
both the optimal quadrupole-quadrupole force and the optimal
one-body spin-orbit force and then ramp up the two pairing strengths
from zero to their optimal values. The results are illustrated in
figure 15, for scenarios in which we separately include isoscalar
pairing, isovector pairing and SU(4) pairing with equal strengths.

As a reminder, the experimental spectrum for $^{48}$Cr shows a
backbend near $I=12$, which as noted earlier is reproduced by our
{\it optimal} hamiltonian. The results of figure 15 make clear that
(a) the backbend cannot be reproduced with pure isoscalar pairing,
but requires isovector pairing as well, and (b) there is no
significant difference between the results obtained with pure
isovector pairing and SU(4) pairing.
\begin{figure}[thp]
\begin{center}
\includegraphics[width=18cm]{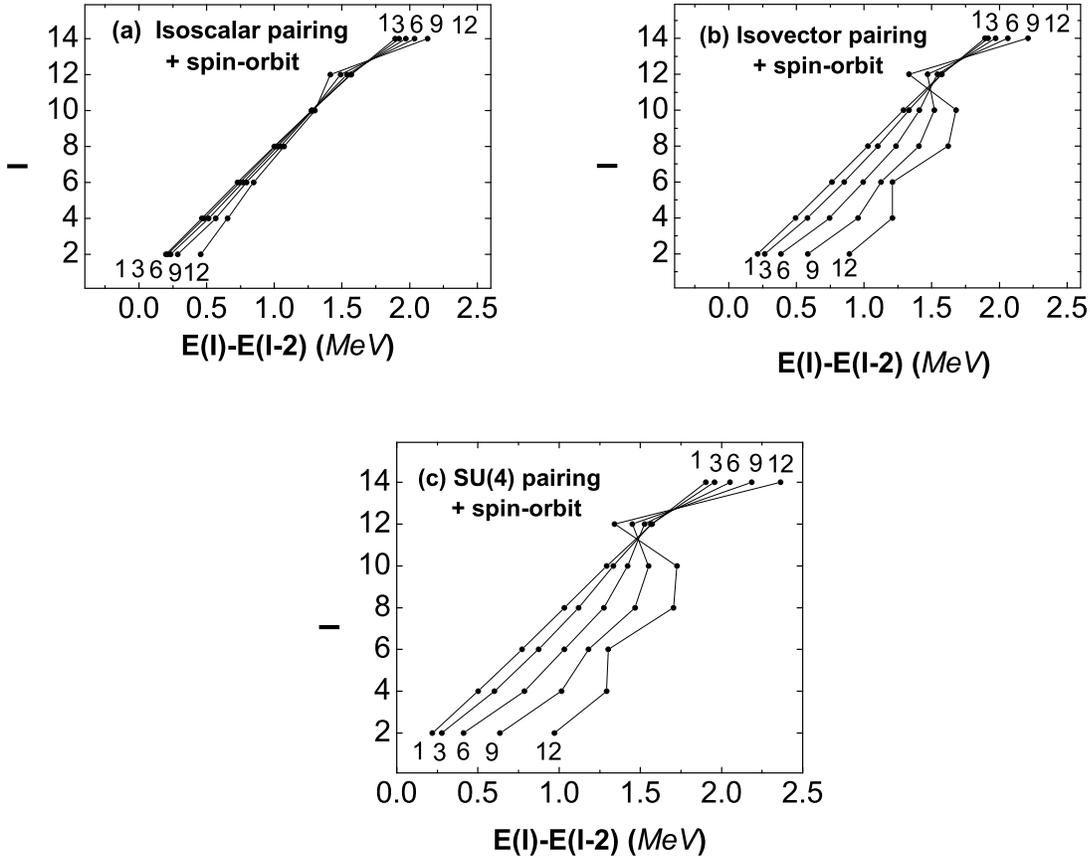}
\end{center}
\vspace{-1cm} \caption{\label{fig15}Calculated energy splittings
$E(I)-E(I-2)$ in $MeV$ in the $^{48}$Cr ground band, for (a)
isovector, (b) isoscalar, and (c) SU(4) pairing, respectively,  as
described in the text. The optimal spin-orbit term is included.}
\end{figure}

The backbend in $^{48}$Cr was discussed extensively in the context
of a shell-model study with a fully realistic hamiltonian in
\cite{Poves}, where it was first shown to derive from isovector
pairing. Our results are in agreement with that earlier conclusion.
To see these points more clearly, we show in figure 16 the numbers
of isovector $S^{\dagger}$ and isoscalar $P^{\dagger}$ pairs as a
function of angular momentum for the optimal hamiltonian.  As in our
treatment of $^{44}$Ti (see Fig. 6), the pair numbers are obtained
by evaluating  $\langle S^{\dagger}\cdot S \rangle $ and $\langle
P^{\dagger}\cdot P \rangle $ and scaling them with respect to the
results that would derive from pure $T=1$ and $T=0$ pairing
hamiltonians, respectively. [Now, however, the analysis is carried
out for a system of four pairs in an $\Omega=10$ shell.] As in ref.
\cite{Poves}, the contribution of isovector pairing in the $J=0^+$
ground state in much larger than the contribution of isoscalar
pairing. As the system cranks to higher angular momenta, the
isovector pairing contribution falls off with angular momentum very
rapidly eventually arriving at a magnitude roughly comparable with
the isoscalar pairing contribution at roughly $J^{\pi}=10^+$.  As
the angular momentum increases even further we see a fairly
substantial increase in the isovector pairing contribution at
$J^{\pi}=12^+$, which according to figure 15 is where the backbend
becomes prominent. After the backbend, both isoscalar and isovector
pairing contributions decrease to near zero as alignment is
achieved.

\begin{figure}[thp]
\begin{center}
\includegraphics[width=16cm]{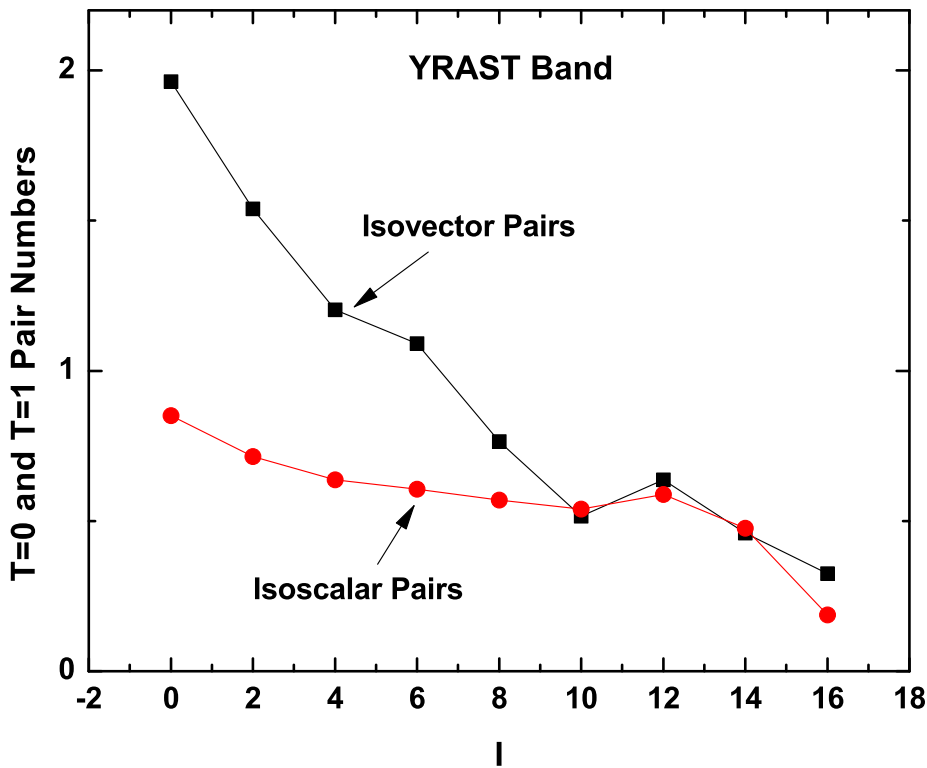}
\end{center}
\vspace{-1cm} \caption{\label{fig16}Calculated numbers of isovector
$S^{\dagger}$ pairs and isoscalar $P^{\dagger}$ pairs in the ground
(YRAST) band of $^{48}$Cr for the optimal values of the hamiltonian
parameters.}
\end{figure}

We have also studied the properties of the lowest excited (YRARE)
band that emerges from the same calculation, a K=2$^+$ band. The
energies of this band are illustrated in figure 17, together with
those of the ground (YRAST) band. From this figure, we conclude that
the backbend in ${48}$Cr does not derive from level crossing.
\begin{figure}[thp]
\begin{center}
\includegraphics[width=16cm]{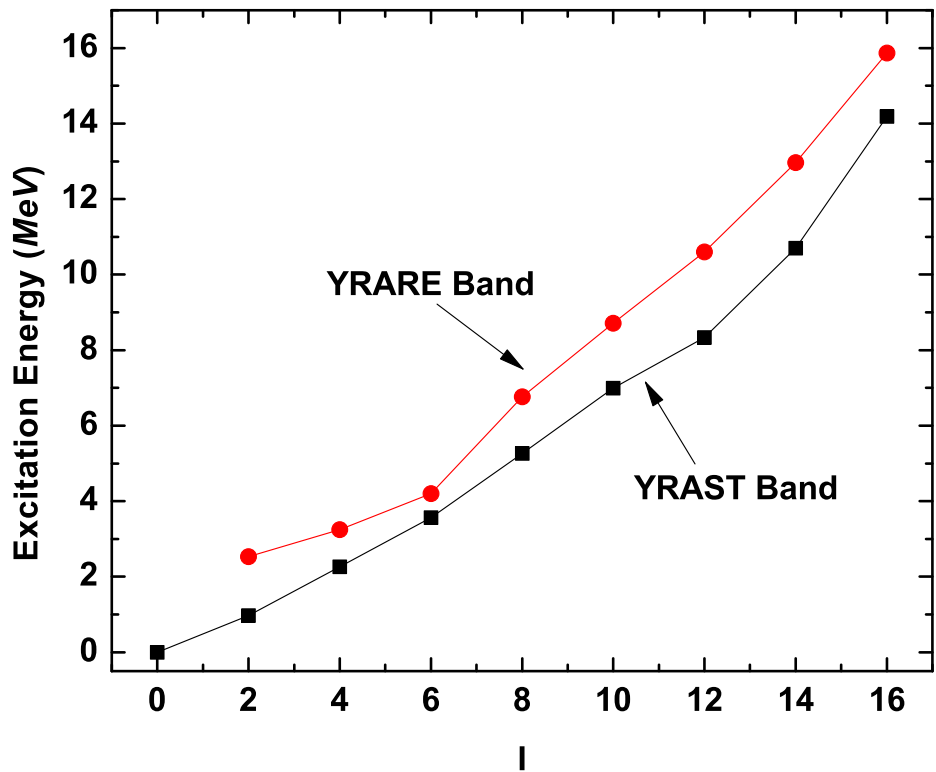}
\end{center}
\vspace{-1cm} \caption{\label{fig17}Calculated excitation energies
of the ground (YRAST) band and the first excited (YRARE) band in
$^{48}$Cr for the optimal values of the hamiltonian parameters. }
\end{figure}

In figure 18, we show the number of isocalar and isovector pairs in
the excited YRARE band, to be compared with the results for the
YRAST band of figure 16. In the backbend region, the numbers of
isoscalar and isovector pairs are found to behave differently in the
two bands. Whereas the numbers of isoscalar and isovector pairs are
roughly the same in the YRAST band (fig. 16), there are
substantially fewer isoscalar pairs than isovector pairs in the
YRARE band (fig. 18). We believe that this is an interesting
observation worthy of further study.

\begin{figure}[thp]
\begin{center}
\includegraphics[width=16cm]{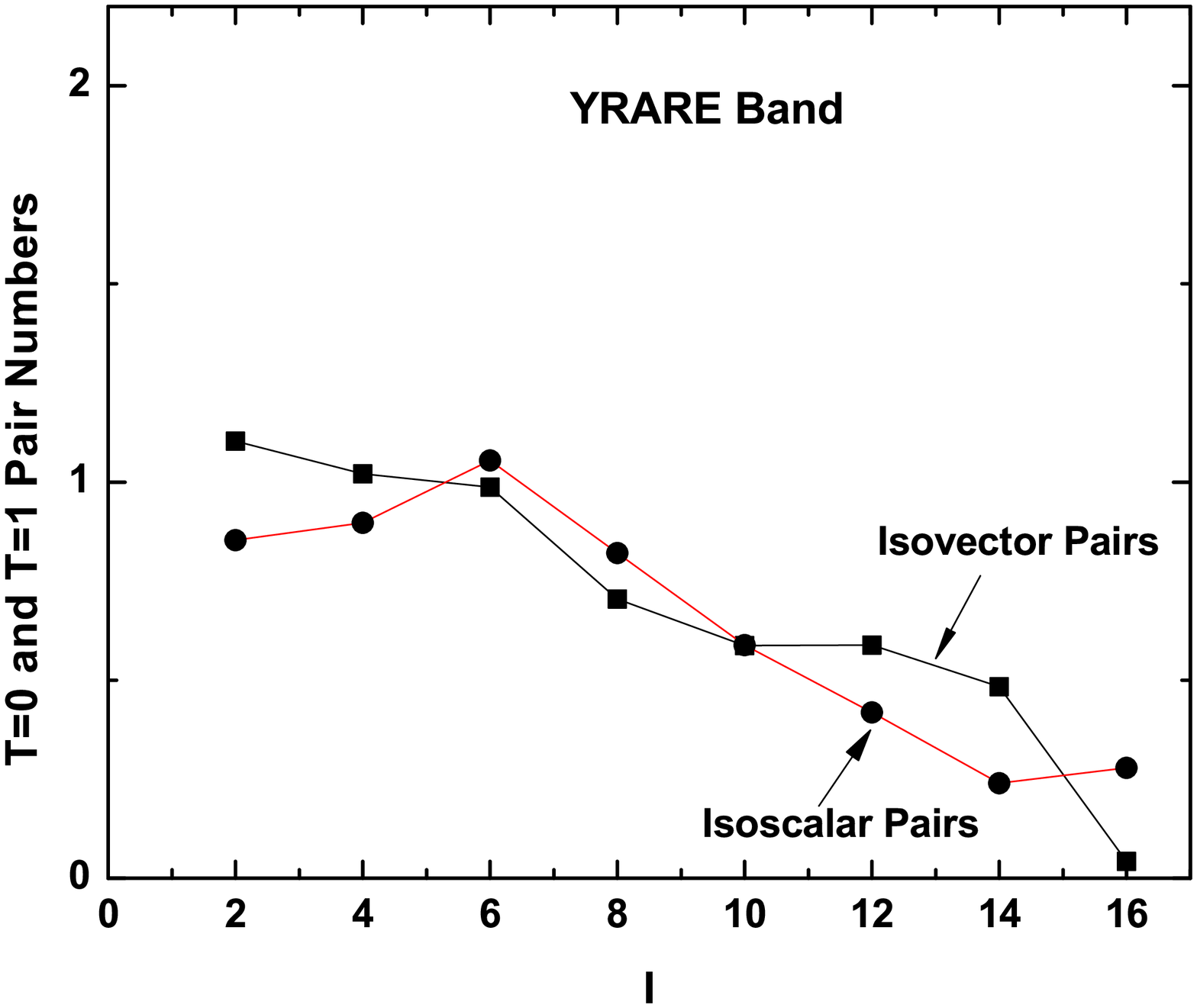}
\end{center}
\vspace{-1cm} \caption{\label{fig18}Calculated numbers of isovector
$S^{\dagger}$ pairs and isoscalar $P^{\dagger}$ pairs in the first
excited (YRARE)  band of $^{48}$Cr for the optimal values of the
hamiltonian parameters.}
\end{figure}

\section{Summary and Concluding Remarks}

In this work, we have reported a systematic shell-model study of
proton-neutron pairing in $2p1f$ shell nuclei using a parametrized
hamiltonian that includes deformation and spin-orbit effects as well
as both isoscalar and isovector pairing. By working in a shell-model
framework we are able to assess the role of the various pairing
modes in the presence of nuclear deformation without violating
symmetries.

We first showed that our parametrized hamiltonian has enough
flexibility to provide a reasonable description of the evolution of
nuclear structure properties in this region. We then probed the role
of the various modes of pairing on deformation with and without a
spin-orbit term. We did this as a function of the number of neutrons
and protons, so as to assess the role both of a neutron excess and
of the number of active particles.

Some of the conclusions that emerged are: (1) in the absence of a
spin-orbit term, isoscalar and isovector pairing have identical
effects at  $N=Z$, but isoscalar pairing ceases to have an
appreciable effect for nuclei with just two excess neutrons; (2) the
non-rotational character of $^{44}$Ti cannot be explained solely in
terms of spin-orbit effects, but requires pairing for its
understanding; (3) in the presence of a spin-orbit interaction,
isoscalar pairing is suppressed even at $N=Z$, (4) the fact that
ground state of $^{46}$V has $J^{\pi}=0^+$ $T=1$  derives primarily
from the spin-orbit interaction and its effect of suppressing
isoscalar pairing, (5) the known backbend in $^{48}$Cr has its
origin in isovector pairing and does not derive from level crossing,
and (6) in the region of the $^{48}$Cr backbend, the numbers of
isoscalar and isovector pairs behave quite differently in the YRAST
and YRARE bands.

\vspace{0.2cm}

{\bf Acknowledgements:} The work reported herein began while two of
the authors, S.P. and N.S., were visiting the Consejo Superior de
Investigaciones Cient\'{i}ficas in Madrid, whose hospitality is
gratefully acknowledged. Much of it was carried out while Y.L. was
visiting the Bartol Research Institute of the University of
Delaware, whose hospitality is likewise acknowledged. The work of
S.P., B.T. and Y.L. was supported in part by the National Science
Foundation under grant \# PHY-0854873, that of N.S. by the Romanian
Ministry of Education and Research through CNCSIS grant Idei nr.
1975, that of A.P. by the projects FPA2009-13377 MICINN(Spain) and
HEPHACOS S2009/ESP-1473 Comunidad de Madrid(Spain), and that of
Y.M.Z. and Y.L. by the National Science Foundation of China under
grant \# 10975096 and by the Chinese Major State Basic Research
Developing Program under grant \# 2007CB815000.


\section*{References}
\begin{thebibliography}{8}
\bibitem{Goodman}
A. L. Goodman, Adv. Nucl. Phys. {\bf 11}, 263 (1979).

\bibitem{DP} J. Dobe\v{s} and S. Pittel, Phys. Rev. {\bf  C57},
688 (1998).

\bibitem{so8} S.C. Pang, Nucl. Phys. {\bf A128}, 497 (1965); J. A.
Evans, G. G. Dussel, E. E. Macqueda and R. P. J. Perazzo, Nucl.
Phys. {\bf A367}, 77 (1981); G. G. Dussel, E. E. Macqueda, R. P. J.
Perazzo and J. A. Evans, Nucl. Phys. {\bf A450}, 164 (1986).
\bibitem{Errea} B. Errea, Ph.D. thesis, Consejo Superior de Investigaciones Cient\'{i}ficas,
unpublished (2009).

\bibitem{Satula} S. Lerma H., B. Errea, J. Dukelsky and W. Satula,
Phys. Rev. Lett. {\bf 99}, 032501 (2007).
\bibitem{Sandulescu} N. Sandulescu, B. Errea, J. Dukelsky,
Phys. Rev.  {\bf C80}, 044335 (2009).
\bibitem{Poves} A. Poves and G. Martinez-Pinedo, Phys. Lett. {\bf B430} 203, (1998).
\bibitem{su3} J. P. Elliott, Proc. R. Soc. London Ser. A {\bf 245}, 128,
562 (1956).

\bibitem{Bertsch} G. F. Bertsch and S. Baroni,  arXiv:nucl-th/0904.2017v2.
\bibitem{Bhatt} K. H. Bhatt and
J. B. McGrory,  Phys. Rev. {\bf C3} 2293 (1971).
\bibitem{quasi-su3}  A. P. Zuker, J. Retamosa, A. Poves, and E. Caurier, Phys. Rev. {\bf C52}, R1741 (1995).
\end{thebibliography}
\end{document}